\documentclass[preprint,amsmath,superscriptaddress,pre]{revtex4}  
\usepackage[dvips]{graphicx}
\usepackage{epsfig}
\usepackage{color}
\usepackage{subfigure}
\usepackage{soul}

\begin{document} 

\title{
Finite-size scaling 
versus dual random variables and
shadow moments
in the size distribution of epidemics
}
\author{\'Alvaro Corral}
\affiliation{%
Centre de Recerca Matem\`atica,
Edifici C, Campus Bellaterra,
E-08193 Barcelona, Spain
}\affiliation{Departament de Matem\`atiques,
Facultat de Ci\`encies,
Universitat Aut\`onoma de Barcelona,
E-08193 Barcelona, Spain
}\affiliation{Barcelona Graduate School of Mathematics, 
Edifici C, Campus Bellaterra,
E-08193 Barcelona, Spain
}\affiliation{Complexity Science Hub Vienna,
Josefst\"adter Stra$\beta$e 39,
1080 Vienna,
Austria
}
\maketitle

In an appealing paper, Cirillo and Taleb \cite{Cirillo_Taleb}
deal with the important issue of the size of epidemic outbreaks, 
understanding size as number of fatalities.
They claim that the distribution of fatalities in historical epidemics seems to be ``extremely fat-tailed'',
which means that the survival function
(or complementary cumulative distribution function) 
decays as power law (pl), asymptotically
(i.e., for very large number of fatalities).
In a formula, 
$S_{\mbox{\tiny pl}}(x) \sim 1/x^{\alpha}$,
where $x$ is the number of fatalities in each epidemic,
$S(x)$ its survival function, 
and $\alpha>0$ the exponent of the survival function,
related to the tail index $\xi$ (measuring fat-tailness)
by $\xi=1/\alpha$. 
Equivalently, 
$f_{\mbox{\tiny pl}}(x) \sim 1/x^{1+\alpha}$,
with $f(x)$ the probability density.

The authors of Ref. \cite{Cirillo_Taleb} correctly mention 
that power laws are characterized by lack of moments;
from a ``physical'' point of view
this means that the moments that do not exist can be considered as taking an infinite value.
In particular, not even the first moment (the mean) is finite if 
$\alpha< 1$.
%
They further state that this lack of moments
for the number of fatalities
is questionable.
%

Of course, the random variable $x$ is bounded by the total world population,
and therefore, in practice, the moments cannot be infinite.
%
%
%
%
Notice that
from the point of view of probability theory there is no fundamental impediment in the fact that the moments become infinite (although the law of large numbers and the central limit theorem do not hold 
when $\alpha$ takes the lowest values and one has to rely on the generalized central limit theorem instead \cite{Bouchaud_Georges}).
However, from the physical perspective, this infinitude can seem embarrassing.

In order to solve the apparent problem of the lack of moments, 
the authors of Ref. \cite{Cirillo_Taleb} 
invent a map from the number of fatalities $x$ to a new random variable $z$
(called the dual variable)
given by
\begin{equation}
z=l - \Delta \ln\left(1-\frac {x-l}{h-l}\right),
\label{dual}
\end{equation}
where $l$ is the lowest value of the variable (1000 for the data of Ref. \cite{Cirillo_Taleb}),
$h$ is the world population,
and $\Delta=h$ or $\Delta=h-l$ with no big difference in the final results as $h\gg l$.
The point of Ref. \cite{Cirillo_Taleb} is that the transformation from $x$ to $z$ is 
innocuous for $x\ll h$, yielding $z\simeq x$,
but ensures that $x=h$ transforms into $z=\infty$
(in practice, $z=x$ even for the largest events on record).

Then the autors claim that the fat tail needs to be studied for the $z$ variable
(which is, in theory, unbounded) instead that for the measured variable $x$. 
So, one fits a fat tail for $z$ and transforms back to get the distribution of $x$,
for which all moments can be easily computed, turning out to be finite
(due to the obvious fact that $\mbox{Prob}[x\le h]=1$, as enforced by the map).  
The moments of $x$ calculated in this way are called the ``shadow moments'' in Ref. \cite{Cirillo_Taleb}.

It is clear that the dual transformation above (\ref{dual}), 
together with the fat-tailness of $z$,
imposes an ad-hoc 
form for the tail of the number of fatalities. 
But there is no justification for such an assumption [Eq. (\ref{dual})], 
and one could equally assume
\begin{equation}
z=\sqrt[k]{l^k - (h^k-l^k) \ln\left[1- \frac{x^k-l^k}{h^k-l^k}\right]},
\label{dualk}
\end{equation}
with $k>0$.
This would yield, for different values of $k$, 
different ad-hoc distributions of the number of fatalities $x$, 
and therefore different values for the corresponding moments.
Other choices are possible, for example,
$$
z=x+h^\gamma\left(\frac 1 {(h-x)^\gamma}-\frac 1 {\Delta^\gamma}\right),
$$
with $\gamma>0$.
Still another arbitrary option would have been to consider that $x$
has a power-law shape but with a sharp truncation \cite{Corral_Serra} at $x=h$,
and calculate the moments of this distribution.

The number of ad-hoc options is infinite, and
each one of them would lead to different values of the (shadow) moments.
Table \ref{table1} confirms this, 
showing the mean of $x$, conditioned to $x>200,000$ fatalities, 
for different assumptions for the map between $x$ and $z$
when $z$ has a power-law tail with exponent $\alpha=0.617$ (value taken from Ref. \cite{Cirillo_Taleb}).

\begin{table}[h]
\begin{center}
\caption{\label{table1} 
Mean value of $x$ conditioned to $x>200,000$
when $z$ has a power-law tail with exponent $\alpha=0.617$ 
\cite{Cirillo_Taleb} for different maps between 
$x$ and $z$, as specified in the first and second columns.
Mean values are calculated from Monte Carlo simulations with
$10^6$ realizations.
trunc pl denotes a power law sharply truncated at $h$.
For the model given by Eq. (\ref{simple}) it can be verified that
the mean scales as $\theta^{1-\alpha}$.
}
\smallskip
\begin{tabular}
{| l|l|r|}
\hline
Model & parameters & $\langle x | x > 200,000\rangle$ \\
\hline 
Eq. (\ref{dual}) \cite{Cirillo_Taleb}& $--$ & $2.57 \times 10^7$\\
\hline
Eq. (\ref{dualk}) & $k=0.1$ & $ 3.8\times 10^6$ \\
id. & $k=0.25$ & $ 1.14\times 10^7$ \\
id. & $k=0.5$ & $ 2.00 \times 10^7$ \\
\hline
trunc pl & $--$ & $ 1.79\times 10^7$ \\ 
\hline
Eq. (\ref{simple}) & $\theta=h/100$ & $2.6 \times 10^6$ \\
id. & $\theta=h/10$ & $ 6.5\times 10^6$ \\
id. & $\theta=h/2$ & $ 1.25 \times 10^7$ \\
id. & $\theta=h$ & $ 1.60 \times  10^7$ \\
\hline
\end{tabular}
\par
\end{center}
\end{table}



\begin{figure}[!ht]
\begin{center}
\includegraphics[width=7.5cm]{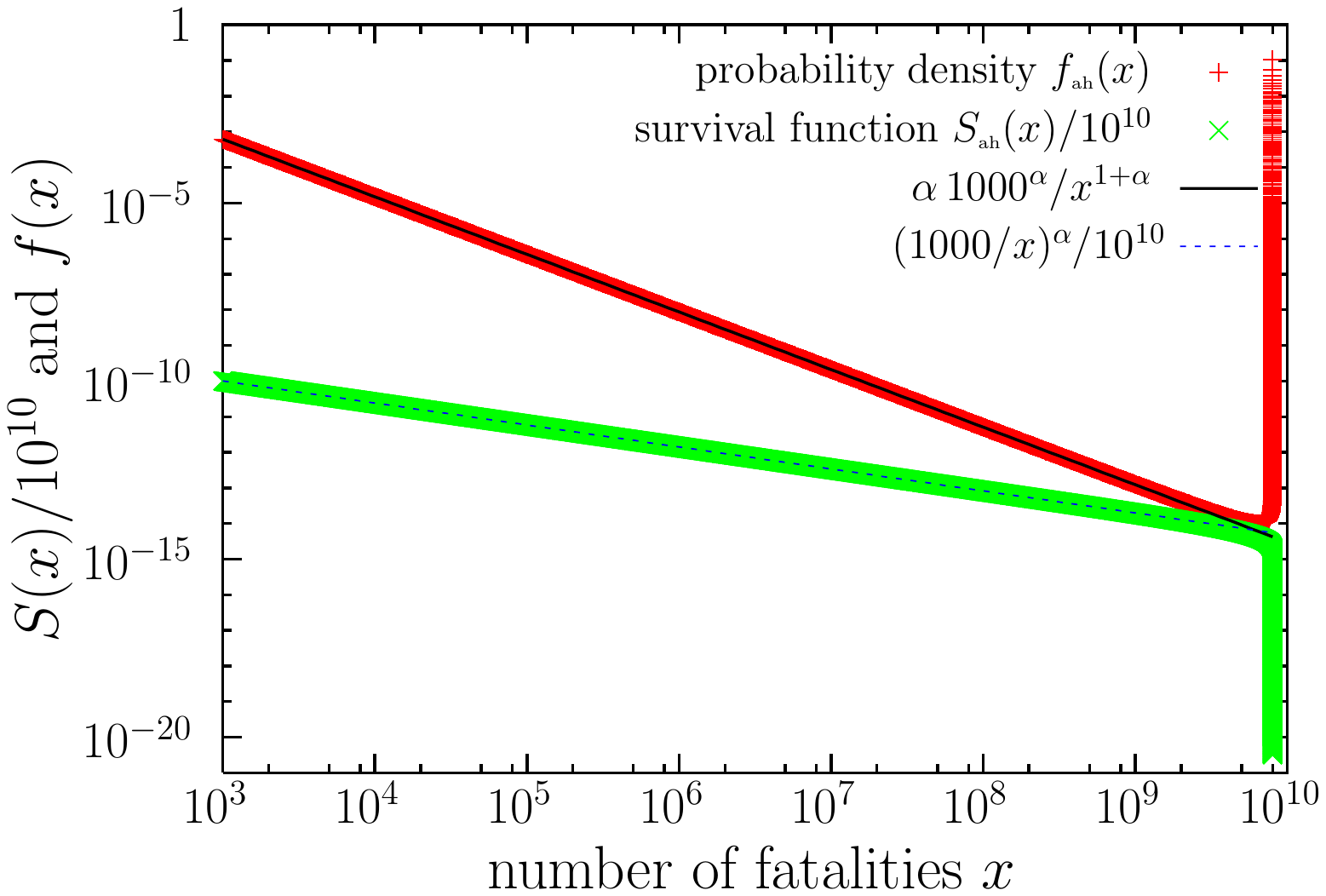}
\end{center}
\caption{
Theoretical
probability density and survival function of the number of epidemic fatalies $x$
obtained from a $z-$variable with a power-law shape for $z\ge 1000$
to which the map of Ref. \cite{Cirillo_Taleb} [Eq. (\ref{dual}) here]  is applied.
The value of the exponent, $\alpha=0.617$, comes from Ref. \cite{Cirillo_Taleb} and $h$ is $8\times 10^9$.
The corresponding power laws are also shown as straight lines.
Notice how the rapid decrease of $S_{\mbox{\tiny ah}}(x)$ at $x\rightarrow h$ translates into a divergence of $f_{\mbox{\tiny ah}}(x)$.
Values of the survival functions are shifted 10 orders of magnitude (factor $10^{10}$) for clarity sake.
}
\label{fig1}
\end{figure}

An illuminating result from the table is that the value of the (conditional) mean for a truncated power law is smaller than the value obtained from the map assumed in Ref. \cite{Cirillo_Taleb}. 
This seems contradictory, as the authors of 
Ref. \cite{Cirillo_Taleb} state that their map provides a smooth truncation
of the distribution.
In fact, the ad-hoc survival function 
resulting from the map in Ref. \cite{Cirillo_Taleb}
(together with a power-law tail for $z$) is
$$
S_{\mbox{\tiny ah}}(x) =S_{\mbox{\tiny pl}}(z(x)) \sim \frac 1 {z(x)^\alpha} =
\frac 1 {[C-\Delta\ln(h-x)]^\alpha},
$$ 
with $C$ just a constant.
Figure \ref{fig1} shows the fast way in which the survival function goes to zero at $x=h$.

The relatively large change in the value of $S(x)$ in a relatively 
small range of $x$
is an indication of a peak in the value of the density $f(x)$
close to $x=h$
(remember that $f(x)=-dS(x)/dx$).
Indeed, one obtains
$$
f_{\mbox{\tiny ah}}(x) \sim 
\frac 1 {(h-x)}
\frac 1 {[C-\Delta\ln(h-x)]^{1+\alpha}},
$$
where the second factor tends to zero when $x\rightarrow h$, 
in the same way as $S_{\mbox{\tiny ah}}(x)$;
however, the first factor goes to infinity, 
and this is the term that dominates,
as shown in Fig. \ref{fig1} (and as one can easily calculate).
This explains the high value obtained for the (conditional) mean
in comparison with a truncated power law, 
as the map in Ref. \cite{Cirillo_Taleb} poses an excess of mass
in the most extreme values of the distribution (tending to a Dirac delta function).
Thus, we conclude that the ad-hoc form of the distribution
resulting from this map is clearly unphysical.
Comparison with a Weibull tail \cite{Coles} seems to indicate
that the distribution belongs to the so-called Weibull maximum domain of attraction, 
with $\xi=-\infty$ (for $x$, not for $z$),
which is the most opposed case to a fat tail.

%
%



Let us set clearly that,
in statistical physics, the ``canonical'' approach for this kind of problems
(when one expects finite-size effects)
is a finite-size scaling (fss) ansatz.
In this framework, one can write for the probability density
$$
f_{\mbox{\tiny fss}}(x)=\frac 1 l \left(\frac l x \right)^{1+\alpha} G\left(\frac x {h^d}\right),
$$
where $d$ is a (positive) scaling exponent
and $G$ an unknown scaling function, 
but whose shape has to be constant for small arguments and decaying fast (for example, exponentially) for large arguments
(the exact formula for $G$ can be rather complicated for very simple models, 
see for instance Ref. \cite{Corral_garcia_moloney_font}).
A simplified (sim), concrete option \cite{Serra_Corral} can be
\begin{equation}
f_{\mbox{\tiny sim}}(x) \propto 
\frac 1 l \left(\frac l x \right)^{1+\alpha} e^{-x/\theta},
\label{simple}
\end{equation}
with $\theta\propto h^d$.
The scaling function does not have to be confused with the slowly varying function
defining fat tails \cite{Cirillo_Taleb}; 
in fact, the scaling function destroys fat-tailness.
Notice that the term ``finite size'' refers to the system size, and not to the random variable $x$ 
(which is a size). 
The idea is that the finite size of the system limits the growth of the size of the phenomenon.
Although the finite-size scaling hypothesis may seem another arbitrary assumption, 
it is supported by an enormous amount
of theory and simulation results \cite{Privman}.

Within this framework one can obtain that the moments of $x$ scale as
$$
\langle x^q \rangle 
\propto 
\left\{
\begin{array}{ll}
{l^{q}} &\mbox{ for } q < \alpha,\\
{l^{\alpha}} {h^{d(q-\alpha)}} &\mbox{ for } q > \alpha,\\
\end{array}
\right.
$$ 
when $\alpha >0$ \cite{Corral_csf}, setting clearly the divergence of the moments for $q>\alpha$
when $h\rightarrow \infty$,
for which fat-tailness is recovered.
Apart from the knowledge of $\alpha$,
the key ingredient to obtain the right scaling of the moments (not their exact expression) 
would be the determination of the exponent $d$.

One could naively assume, from wrong dimensional analysis, that $d=1$,
but this is not the case in many well studied models \cite{Corral_garcia_moloney_font}.
If one attemps an empirical estimation,
given diverse isolated regions of the world 
(each one with its total population $h$) one could study the scaling behavior
of the moments or of the distribution with $h$.
However, due to the current scarcity of data for epidemic fatalities, 
the only option is the use of computer simulations, 
with a model in the same universality class than real epidemics.
Of course, this is also unknown, nowadays, 
and so, the real moments of the distribution will remain unknown as well.

Summarizing,
fat-tailed distributions are characterized by diverging moments.
This implies that, for a (finite) sample, the empirical moments 
grow with the number of data
(and in particular the law of large numbers does not hold for $\alpha<1$).
One could assume a fat tail to model the empirical data available for the size of epidemics, although one must expect a deviation from a power law for
very large values of the size $x$.
Under the finite-size scaling hypothesis this deviation is given by the scaling function $G$ and starts taking place at a scale $\propto h^d$.
Alternatively, one could assume ad-hoc expressions for this deviation, 
as in Ref. \cite{Cirillo_Taleb}.
In any case, the scale at which the deviation takes place 
(whatever its mathematical form)
seems to be
larger than the largest values of $x$ observed, 
and therefore are unobservable, in practice.

If the deviations from power law behavior are unobservable, 
the empirical moments are not influenced by these deviations, 
and therefore, do not converge to the (unknown) theoretical values,
for the number of empirical data available.
In other words, empirical moments are ill-defined.
The lesson learned from here is that, as moments are ill-defined, 
it is pointless to try to calculate them and one should rely on other metrics 
to assess risk.

I acknowledge my colleague Isabel Serra for discussions
and
support from projects
FIS2015-71851-P and
PGC-FIS2018-099629-B-I00
from Spanish MINECO and MICINN.

\end{document}